# Markov random walk under constraint for discovering overlapping communities in complex networks

**Di Jin[1], Bo Yang[1], Carlos Baquero[2], Dayou Liu[1], Dongxiao He[1] and Jie Liu[1]**

[1] College of Computer Science and Technology, Jilin University, Changchun 130012, China

[2] CCTD/DI, University of Minho, Braga, Portugal

E-mail: jindi.jlu@gmail.com, ybo@jlu.edu.cn, cbm@di.uminho.pt, liudy@jlu.edu.cn, hedongxiaojlu@gmail.com and liu_jie@jlu.edu.cn

**Abstract.** Detection of overlapping communities in complex networks has motivated recent research in the relevant fields. Aiming this problem, we propose a Markov dynamics based algorithm, called UEOC, which means, "unfold and extract overlapping communities". In UEOC, when identifying each natural community that overlaps, a Markov random walk method combined with a constraint strategy, which is based on the corresponding annealed network (degree conserving random network), is performed to unfold the community. Then, a cutoff criterion with the aid of a local community function, called conductance, which can be thought of as the ratio between the number of edges inside the community and those leaving it, is presented to extract this emerged community from the entire network. The UEOC algorithm depends on only one parameter whose value can be easily set, and it requires no prior knowledge on the hidden community structures. The proposed UEOC has been evaluated both on synthetic benchmarks and on some real-world networks, and was compared with a set of competing algorithms. Experimental result has shown that UEOC is highly effective and efficient for discovering overlapping communities.

## 1. Introduction

Many complex systems in real world exist in the form of networks, such as social networks,

biological networks, Web networks, etc., which are collectively referred to as complex networks. One of the main problems in the study of complex networks is the detection of community structure, i.e. the division of a network into groups of nodes having dense intra-connections, and sparse inter-connections [1]. In the last few years, many different approaches have been proposed to uncover community structure in networks; the interested reader can consult the excellent and comprehensive survey by Fortunato [2].

However, it is well known that many real-world networks consist of communities that overlap because nodes are members of more than one community [3]. Such observation is visible in the numerous communities each of us belongs to, including those related to our scientific activities or personal life (school, hobby, family, and so on). Another, biological example is that a large fraction of proteins simultaneously belong to several protein complexes [4]. Thus, hard clustering is inadequate for the investigation of real-world networks with such overlapping communities. Instead, one requires methods that allow nodes to be members of more than one community in the network.

Recently, a number of approaches to the detection of overlapping communities in graphs have been proposed. One type, among these methods, is based on the idea of clique percolation theory, i.e. that a cluster can be interpreted as the union of small fully connected sub-graphs that share nodes [3, 5, 6]. Another type of method transforms the interaction graph into the corresponding line graph in which edges represent nodes and nodes represent edges, and then apply a known clustering algorithm on the line graph [7-10]. A third type of method, which our work belongs to, discovers each natural community that overlaps by using some local property based approach [11-13].

Though there are already some algorithms for detecting overlapping communities, which have been presented recently, there is still ground for improvements in the performance. Moreover, while the Markov dynamics model has been successfully used in the area for community detection [14-18], unfortunately, the existing approaches still largely lack the ability to deal with overlapping communities. To the best of our knowledge, only to recent works, available in pre-print format, constitute exceptions to this. Esquivel et al. [19] generalized the map equation method [16] by releasing the constraint that a node can only belong to one module codebook, and made it able to measure how well one can compress a description of flow in the network when we partition it into modules with possible overlaps. Kim et al. [20] extended the map equation method [16], which is originally developed for node communities, in order to find link communities in networks, and made it able to find overlapping communities of nodes by partitioning links instead of nodes.

In this paper, an algorithm UEOC based on the Markov dynamics model is proposed to discover communities that share nodes. In the UEOC approach, so as to detect all the natural communities, a Markov random walk method is combined with a new constraint strategy, which is based on the corresponding annealed network [21], and used to unfold each community. Then, a cutoff criterion with the aid of conductance, which is a local community fitness function [22], is applied to extract the emerged community. These extracted communities will naturally overlap if this configuration is present in the network. Furthermore, a strong point in our approach is that UEOC is not sensitive to the choice of its only parameter, and needs no prior knowledge on the community structure, such as the number of communities.

## 2. Algorithm

*2.1. Overview of the UEOC*

Let $N = (V, E)$ denote an unweighted and undirected network, where $V$ is the set of nodes (or vertices) and $E$ is the set of edges (or links). We define a cover of network $N$ as a set of overlapping communities with a high density of edges. In this case, some nodes may belong to more than one community. This is an extension of the traditional concept of community detection (in which each node belongs to a single community), to account for possible overlapping communities. In our case, detecting a cover amounts to discovering the natural community of each node in network $N$.

The straightforward way to compute a cover of a targeted network is to repeatedly detect the community for each single node. This is, however, computationally expensive. The natural communities of many nodes often coincide, so most of the computer time is spent to rediscover the same modules over and over. For our algorithm, UEOC, a more efficient way is summarized as follows.

S1. Pick node $s$ with maximum degree, which has not been assigned to any community;

S2. Unfold the natural community of node $s$ by a Markov random walk method combined with a constraint strategy;

S3. Extract the emerged community of node $s$ by a cutoff criterion based on a conductance function;

S4. If there are still nodes that have not been assigned to any community, repeat from S1.

The core of UEOC is how to unfold and extract the natural community of each node, and this directly decides the performance of our algorithm. For the first goal, a Markov random walk method combined with a constraint strategy is proposed here, and this will result in making each community clearly visible. For the second one, a cutoff criterion based on the conductance function is presented, so as to precisely extract the emerged community.

*2.2. Unfolding a community*

Given a network $N = (V, E)$, consider a stochastic process defined on $N$, in which an imaginary agent freely walks from one node to another along the links between them. When the agent arrives at one node, it will randomly select one of its neighbors and move there.

Assume that $X = \{X_t, t \geq 0\}$ denote the agent positions, and $P\{X_t = j, 1 \leq j \leq n\}$ denote the probability that the agent arrives at node $j$ after $t$ steps walking. For $t > 0$ we have $P\{X_t | X_0, X_1, …, X_{t-1}\} = P\{X_t | X_{t-1}\}$. That is, the next state of the agent is completely decided by its previous state, which is called a Markov property. So, this stochastic process is a discrete Markov chain and its state space is $V$. Furthermore, $X_t$ is homogeneous because $P\{X_t = j | X_{t-1} = i\} = p_{ij}$, where $p_{ij}$ is the transition probability from node $i$ to node $j$. In terms of the adjacency matrix of $N$, $A = (a_{ij})_{n \times n}$, $p_{ij}$ is defined as (1).

$$p_{ij} = \frac{a_{ij}}{\sum_r a_{ir}} \tag{1}$$

Lets consider the Markov dynamics model above. Given a specific source node $s$ for the agent, let $\alpha_s^l(i)$ denotes the probability that this agent starts from node $s$ and eventually arrives at an arbitrary destination node $i$ within $l$ steps. The value of $\alpha_s^l(i)$ can be estimated iteratively by (2).

$$\alpha_s^l(i) = \sum_{r=1}^{n} \alpha_s^{l-1}(r) \cdot p_{ri} \quad (2)$$

Here, $\alpha_s^l$ is called the *l* step transition probability distribution (vector). Note that the sum of the probability values arriving at all the nodes from source node *s* will be 1, i.e. $\sum_{i=1}^{n} \alpha_s^l(i) = 1$. When step number *l* equals to 0, which means the agent is still on node *s*, then $\alpha_s^0(s)$ equals to 1 and $\alpha_s^0(i)$ equals to 0 for each $i \neq s$.

As the link density within a community is, in general, much higher than that between communities, a random walk agent that starts from the source node *s* should have more paths, to choose from, to reach the nodes in its own community within *l* steps, when the value of *l* is suitable. On the contrary, the agent should have much lower probability to arrive at the nodes outside its associated community. In other words, it will be difficult for the agent to escape from its existing community by passing those "bottleneck" links and to arrive at other communities. Thus, in general, vector $\alpha_s^l$ should broadly meet the condition (3) when the step number *l* is suitable. In this equation, $C_s$ denotes the community where node *s* is situated.

$$\forall_{i \in C_s} \forall_{j \notin C_s} : \alpha_s^l(i) > \alpha_s^l(j) \quad (3)$$

However, though the above Markov method is well suitable for some simple networks, such as the benchmark graphs in the Newman model [1] and some small real networks, it is not so effective for some complicated networks, like the benchmark graphs in the Lancichinetti model [23] (such as in the example of figure 1(a)) and some large-scale real networks. Furthermore, this method is very sensitive to the choice of the step number *l*, which depicts a crucial influence in its performance.

There is an example, depicting vector $\alpha_s^l$ and shown in figure 1(b). As we can see, the associated probability values of many within community nodes are smaller than that of the outside community nodes, thus it cannot unfold a clear community. This also means that $\alpha_s^l$ does not fit condition (3) well enough for this relatively complicated network. Moreover, the result shown in this figure is the best performance case (when *l* equals to 3). It will become worse when step number *l* is greater or smaller than 3.

In order to overcome these drawbacks, a Markov random walk method combined with a constraint strategy based on the corresponding annealed network is proposed here. The idea of our method arises in the intuition that a Markov random process on a network with community structure is different from that process on its corresponding annealed network without communities. Considering this, in each step, the probability that an agent starts from a specific source node *s* and arrives at each destination node *i* will be defined as the difference between its associated probability computed on the community network *N* and that on the corresponding annealed network *R*. Due to *R* having no community structure, the link density within a community in network *N* should be much higher than that in *R*, while the link density between communities in *N* should be much lower than that in *R*. Thus, under the constraint brought by the annealed network, this agent will be deterred from escaping it's associated community and reach the nodes outside that community. This will also cause that, the computed probability value of each within community node will be high, whilst that of each outside node will be relatively low and in most cases even equal to 0.

Given network $N = (V, E)$ with its degree distribution $D$, and the corresponding annealed network $R = (V', E')$ with its degree distribution $D'$, it should be so that $V = V'$, $D = D'$ while $E \neq E'$. This means that $N$ and $R$ have the same degree distribution [21]. Let $A = (a_{ij})_{n \times n}$ denote the adjacency matrix of network $N$. There will be $D = diag(d_1, \ldots d_n)$, in which $d_i = \sum_j a_{ij}$ denotes the degree of node $i$. Assume that $B = (b_{ij})_{n \times n}$ is the adjacency matrix (also called probability matrix) of $R$. We have $b_{ij} = d_i d_j / \sum_{r=1}^{n} d_r$, which denotes the expected number of links (or called expected link probability) between nodes $i$ and $j$. Let $q_{ij}$ denote the transition probability from node $i$ to node $j$ on graph $R$. It will be defined as (4).

$$q_{ij} = \frac{b_{ij}}{\sum_r b_{ir}} \quad (4)$$

Considering the constraint generated by this annealed network $R$, let $\beta_s^l(i)$ denote the probability that this agent starts from the source node $s$ and eventually arrives at an arbitrary destination node $i$ within $l$ steps. The value of $\beta_s^l(i)$ can be estimated iteratively by (5).

$$\beta'^l_s(i) = \max\left(\sum_{r=1}^{n} \beta_s^{l-1}(r) \cdot p_{ri} - \sum_{r=1}^{n} \beta_s^{l-1}(r) \cdot q_{ri}, 0\right)$$
$$\beta_s^l(i) = \frac{\beta'^l_s(i)}{\sum_{r=1}^{n} \beta'^l_s(r)} \quad (5)$$

It's obvious that $\sum_{r=1}^{n} \beta_s^{l-1}(r) \cdot p_{ri}$ denotes the transition probability from node $s$ to node $i$ within $l$ steps on network $N$, while $\sum_{r=1}^{n} \beta_s^{l-1}(r) \cdot q_{ri}$ denotes that probability computed on annealed network $R$. Furthermore, we make each $\beta_s^l(i)$ always a nonnegative value. Since the sum of the probability values arriving at all the nodes from source node $s$ should be 1, we also normalize $\beta_s^l(i)$ after each step.

An example for the proposed $l$ step transition probability distribution $\beta_s^l$ is shown in figure 1(c). As we can see, the associated probability values of almost all within community nodes are greater than that of the outside community nodes, except for only a few special nodes whose community relations may be not very clear. In particular, there are 766 out of 903 outside community nodes whose probability values are 0. Thus, it's obvious that, $\beta_s^l$ can meet condition (3) well, and unfold a very clear community for source node $s$. Moreover, this Markov process will convergence very well when step number $l$ is greater than some value (in particular, 20 was found to be a good lower value). Thus, its performance is not as sensitive as before to the parameter $l$. Later, we will offer some detailed analysis on this parameter.

Furthermore, most complex networks have power-law degree distribution, which means there are more paths arriving at the nodes with high degrees than those with low degrees. If not compensated, this will lead to detrimental effects when unfolding communities. Thus, we take into account the effect of power-law degree distribution in complex networks, and propose a further improved $l$ step transition probability distribution $\psi_s^l$ as defined as (6), where $d_i$ denotes the degree of node $i$. Note that this equation is not iteratively computed.

$$\psi'^l_s(i) = \frac{\beta_s^l(i)}{d_i}, \quad \psi_s^l(i) = \frac{\psi'^l_s(i)}{\sum_{r=1}^n \psi'^l_s(r)} \tag{6}$$

There is an example for vector $\psi_s^l$ shown as figure 1(d). As we can see, this improved method can unfold a more clear community, and $\psi_s^l$ can meet (3) a little better than $\beta_s^l$.

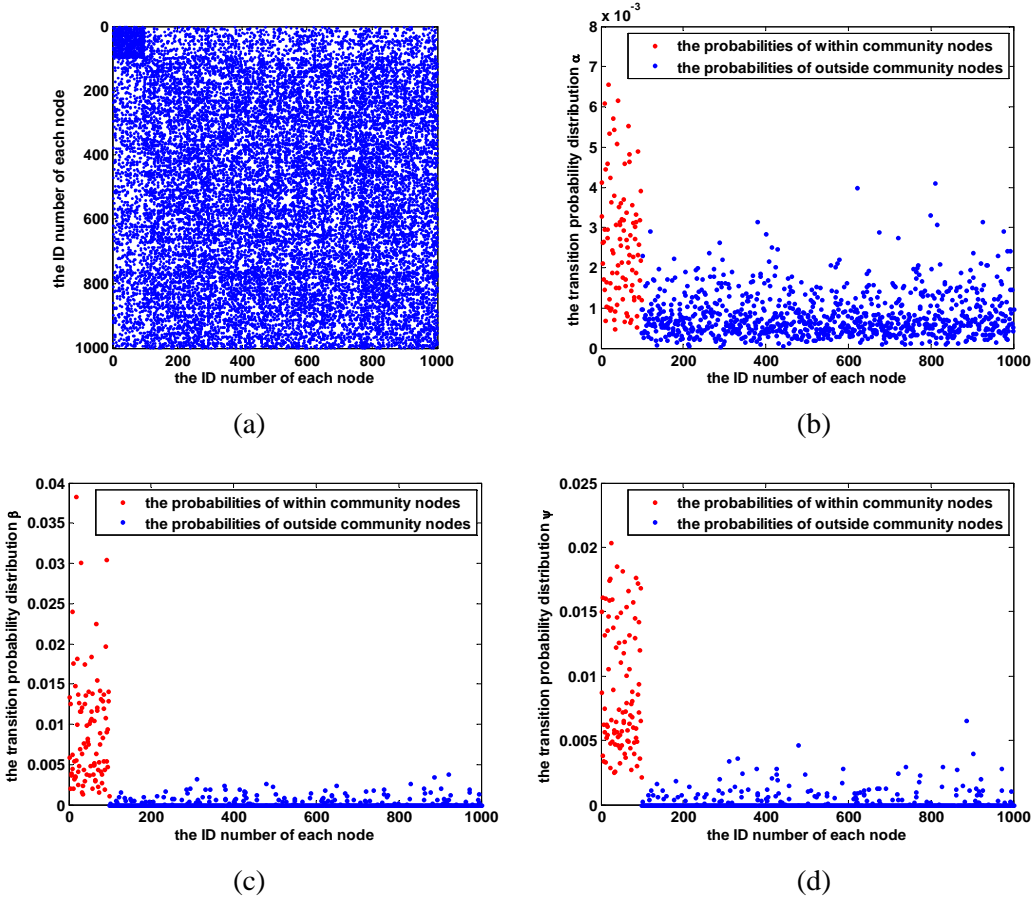

(a)  (b)  (c)  (d)

**Figure 1.** An illustration for the Markov random walk method combined with our constraint strategy to unfold a community. **(a)** A benchmark network with power-law distribution of degree and community size by Lancichinetti model [23]. In this network, the number of nodes is 1000, the minimum community size is 20, the mixing parameter $\mu$ is 0.3, and the number of overlapping nodes is 400. Here we just consider the largest community, which contains 97 nodes. They are all put on the top of the nodes sequence. The selected source node $s$ is the one with maximum degree in this community. **(b)** The generated $l$ step transition probability vector $\alpha$ according to (2). The best performance appears when $l$ equals 3. **(c)** The generated $l$ step transition probability vector $\beta$ according to (5). Its performance is insensitive to the increase of parameter $l$ after it is greater than 20. **(d)** The generated $l$ step transition probability vector $\psi$ according to (6). Its performance is insensitive to the increase of parameter $l$ after it is greater than 20.

Based on this above idea, a method for unfolding the community, which contains a specific source node *s*, will be described bellow. This method is called UC (Unfolding Community).

S1. Calculate the *l* step transition probability vector of node *s*, which is $\psi_s^l$;

S2. Rank all the nodes according to their associated probability values in descending order, producing the sorted node list *L*.

After these two steps, almost all the within community nodes will be ranked on the top of the sorted node sequence *L*. The target community has now clearly emerged and is ready for detection. Now, by properly setting a cutoff point (to be explained in the next section), we can precisely extract the community's nodes.

**Proposition 1.** The time complexity of the UC is $O(ln^2)$, where *l* is the step number of the random walk agent, and *n* corresponds to the number of nodes in network *N*.

***Proof.*** In S1, It's obvious that the time to compute $\psi_s^l$ by (5) and (6) will be $O(ln^2)$. In S2, the time to rank $\psi_s^l$ will be $O(n\log n)$, derived from quick sorting algorithms. Thus, the time complexity of the UC will be $O(ln^2)$.

*2.3. Underlying mechanism in Unfolding Community*

Let an agent freely walk on a network *N*. For any node *i*, *j* and step number *l*, the probability that the agent starts from node *i* and arrives at node *j* within *l* steps will be $\alpha_i^l(j)$ according to (2). This is a discrete Markov process.

From the analysis of [24], according to large deviation theory, we know that the Markov chain has *n* local mixing states, and the *i-th* local mixing state corresponds to the number of communities *i*. In particular, if the network *N* has a well-defined community structure with *k* clear communities, this Markov chain will stay, in a stable way, in the *k-th* local mixing state during a period of time with a probability of 1, called a metastable state in this situation.

According to this theory [24], all the local mixing times of the Markov chain can be estimated by using the spectrum of its Markov generator (normalized graph Laplacians) $M = I - P$, where *I* is the identity matrix and *P* is $(p_{ij})_{n \times n}$ according to (1). For an undirected network, *M* is positive semi-definite and has *n* non-negative real-valued eigenvalues ($0 = \lambda_1 \leq \lambda_2 \leq \cdots \leq \lambda_n \leq 2$). Let $T_i^{ent}$ and $T_i^{ext}$ be the entering time and exiting time of the *i-th* local mixing state. We have $T_i^{ext} = \frac{1}{\lambda_i}(1 + o(1))$. Reasonably, we can also use the exiting time of the (*i*+1)-*th* local mixing state to estimate the entering time of the *i-th* local mixing state. That is $T_i^{ent} = T_{i+1}^{ext} = 1/\lambda_{i+1}$.

There is an example shown as figure 2. In order to depict this Markov process more clearly, we use a simple network and adopt the transition probability matrix instead of the transition probability vector here. Figure 2(a) shows a Newman network [1] which contains a known community structure with four clear communities. Figure 2(b) shows the spectrum of this network. Figure 2(c) shows the exiting time of each local mixing state. Especially, it also offers the entering time and exiting time of the 4-*th* local mixing state, which corresponds to a metastable state and evidences the real community structure of the network. Figure 2(d)-(f) depict the process of the Markov chain when it goes through the metastable state and finally reaches the global mixing state. Figure 2(d) denotes the $t_1$ ($t_1 = 2 \approx T_4^{ent} = T_5^{ext} = 1/\lambda_5$) step transition probability matrix when it begins to enter the metastable state which corresponds to the four clear communities. Figure 2(e) denotes the $t_2$ ($t_2 = 7 \approx T_4^{ext} = 1/\lambda_4$) step transition probability matrix when it begins to exit this metastable state. Figure 2(f) shows the $t_3$ ($t_3 = 20 > T_1^{ent} = T_2^{ext} = 1/\lambda_2$) step transition probability matrix after it enters the global mixing state and eventually converges to this state, where we now have $\alpha_i^{t_3}(j) = d_j / \sum_r d_r$ for each node $i$ and $j$.

As we can see from this example, if a network has a community structure with $k$ clear communities, there will be a short entering time ($1/\lambda_{k+1}$) and a relatively long exiting time ($1/\lambda_k$) for the $k$-th local mixing state, which denotes a metastable state. During this period of time, each community locally mixes together and we can observe the $k$ communities of the network. Furthermore, the Markov chain will quickly converge to the global mixing state after it enters this state ($> 1/\lambda_2$).

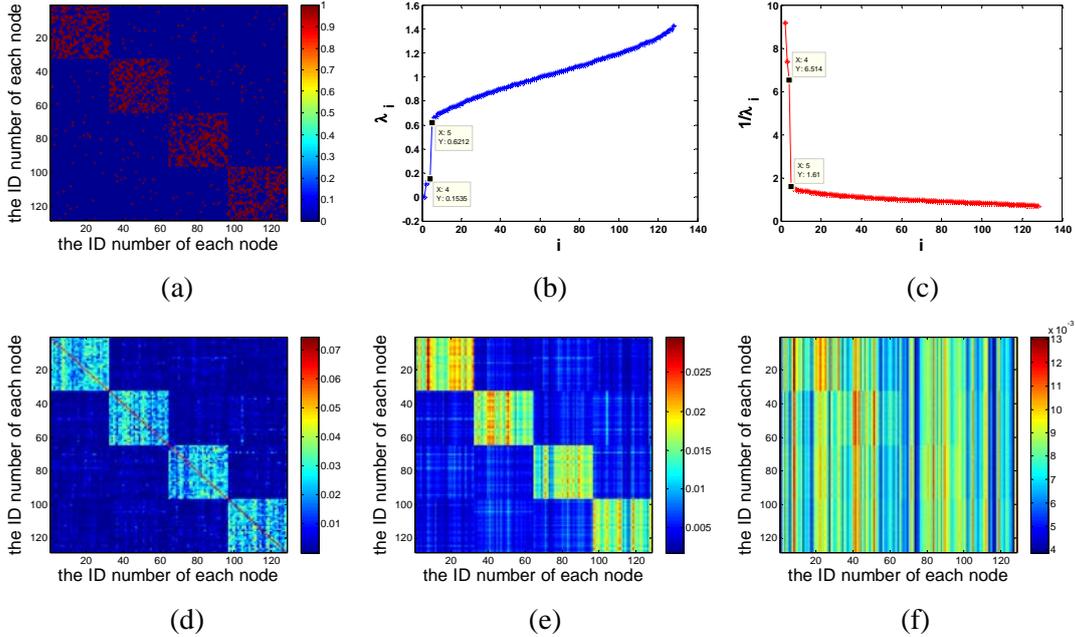

**Figure 2.** An example to demonstrate the characteristic of the iteration process for the Markov chain according to (2) on a benchmark network by Newman model [1]. **(a)** A Newman network consists of 128 nodes divided into four groups of 32 nodes. Each node has on average 14 edges connecting it to members of the same group and 2 edges to members of other groups, with the expected degree 16. **(b)** The spectral distribution ($\lambda_i$) of this network. **(c)** The exiting time ($1/\lambda_i$) of the local mixing state for each number of communities. **(d)** The $t_1$ ($t_1 = 1/\lambda_5$) step transition probability matrix when the Markov chain begins to enter the metastable state. **(e)** The $t_2$ ($t_2 = 1/\lambda_4$) step transition probability matrix when the Markov chain begins to exit the metastable state. **(f)** The $t_3$ ($t_2 > 1/\lambda_2$) step transition probability matrix when the global stable state is finally reached.

Though the Markov chain of a random walk on the network contains a metastable state that corresponds to its real community structure, it will eventually reach the global mixing state, which denotes a trivial solution. For detecting communities, the followed intuition is that, if we can make the Markov chain remain and converge to this metastable state by deliberately adjusting transition probabilities, a nontrivial solution, which corresponds to the real community structure of the network, will be naturally attained due to this adjustment.

Starting from this intuition, here we consider the difference between the Markov random process on a network $N$ with community structure and that on its corresponding annealed network $R$ without communities. It's obvious that, the link density within a community in network $N$ will be much higher than that in $R$, while the link density between communities in $N$ will be much lower than that in $R$. Then, for any node $i, j$ and step number $l$, if the $l$ step transition probability from node $i$ to node $j$ computed on the community network $N$ is no better (lower) than that computed on the annealed network $R$, we will have reason to believe that node $i$ will have no chance of being in a same community of node $j$. Thus, we adjust and rescale the transition probability according to (5), which will make the probability that the constrained walker (agent) starts from node $i$ and arrive at node $j$, within $l$ steps, be null. It's obvious that this will make the agent have almost no chance escaping from its own community, and the constrained Markov chain can hardly exit the metastable state that corresponds to the real community structure of the network.

There is an example shown as figure 3, which corresponds to the case in figure 2. We find out that this constrained Markov chain can also enter the metastable state corresponding to the real community structure at time $t_1 = 1/\lambda_5$, which is shown as figure 3(a). However, it will not begin to exit the metastable state at time $t_2 = 1/\lambda_4$, but has almost converged to this state, which is shown as figure 3(b). At time $t_2 > 1/\lambda_2$, it has completely converged to the metastable state, which depicts the four clear communities shown as figure 3(c).

As we can see from this example, the constrained Markov chain (according to (5)) has similar characteristics to the iteration process with the unconstrained Markov chain (according to (2)). However, the difference is that the unconstrained Markov chain will quickly converge to the global mixing sate after it enters this state ($> 1/\lambda_2$), while the constrained Markov chain will quickly converge to the metastable sate that shows its $k$ real communities after it enters this state ($> 1/\lambda_{k+1}$). Therefore, we can make use of the spectrum of a network to approximate the convergence characteristics of the method UC. In fact, if the network has $k$ real communities, the convergence time of the UC should be only a little longer than the entering time of the $k$-th local mixing state ($> 1/\lambda_{k+1}$).

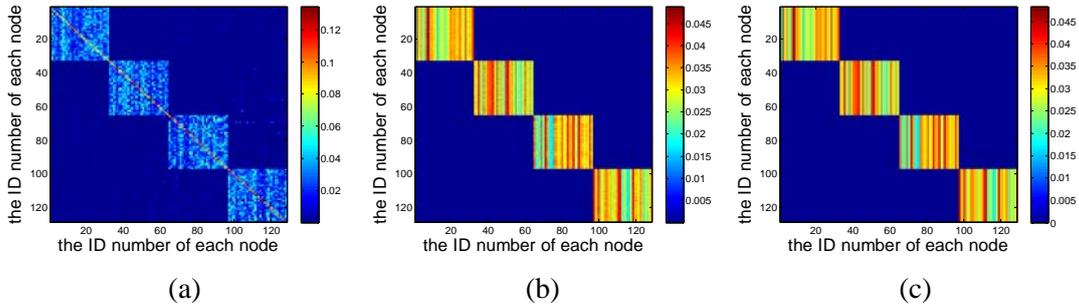

(a)          (b)          (c)

**Figure 3.** An example that demonstrates the characteristics of the iteration process for the constrained Markov chain according to (5) on the network which is shown as figure 2(a). **(a)**

The $t_1$ ($t_1 = 1/\lambda_5$) step transition probability matrix of the constrained Markov chain, which corresponds to figure 2(d). **(b)** The $t_2$ ($t_2 = 1/\lambda_4$) step transition probability matrix of the constrained Markov chain, which corresponds to figure 2(e). **(c)** The $t_3$ ($t_2 > 1/\lambda_2$) step transition probability matrix of the constrained Markov chain, which corresponds to figure 2(f).

It is noteworthy that the above analysis on the characteristics of the iteration process of our UC method is in a close to ideal situation. It becomes more complex when dealing with networks depicting overlapping communities and some more complex topological properties, such as the sample network in figure 1(a) and for some large real-world networks. However, as we can see from figure 1, our UC method can still clearly unfold each node's community and it's also effective in this relatively complicated situation. Later, we will give some detailed analysis on the convergence characteristics of the UC in the experimental section of this article.

*2.4. Extracting the emerged community*

As the community emerged from UC by computing the $l$ step transition probability vector $\psi_s^l$, so that the associated probability values of the within community nodes are much greater than those of outside community nodes, the community associated to source node $s$ can be easily distilled by designing a suitable cutoff criterion.

At first we considered a simple method that takes the average probability value $\varepsilon$ of all the nodes as a cutoff value, which is defined as $\varepsilon = \sum_{i=1}^{n} \psi_s^l(i)/n$. Then we extract the nodes whose associated probabilities are greater than $\varepsilon$ as the detected community. For instance, in figure 4(b), the black line denotes the average probability $\varepsilon$ as cutoff value. We find out that, although this can make all the within community nodes present in the same community, it will also include several outside nodes into this community. Let the structural similarity [25] of two arbitrary sets $v$ and $w$ be $|v \cap w|/\sqrt{|v| \cdot |w|}$. This average cutoff method will show that the structural similarity between the extracted community and the actual community is 0.8068. It's obvious that the method is far from ideal to extract the emerged community on the sample network, even though this network already depicts some complexity.

In order to improve this result and propose a more effective cutoff method, a well-known conductance function [22], corresponding to the weak definition of community [26], is used here in substitute of the less efficient $\varepsilon$ boundary.

The conductance can be simply thought of as the ratio between the number of edges inside the community and those leaving it. More formally, conductance $\phi(S)$ of a set of nodes $S$ is $\phi(S) = c_S/\min(\text{Vol}(S), \text{Vol}(V \setminus S))$, where $c_S$ denotes the size of the edge boundary, $c_S = |\{(u, v) : u \in S, v \notin S\}|$, and $\text{Vol}(S) = \sum_{u \in S} d_u$, where $d_u$ is the degree of node $u$. Thus, in particular, more community-like sets of nodes have lower conductance. Moreover, this community function has some local characteristics, making it suitable to extract the emerged community.

Based on the ranked node list **L** obtained from UC, the emerged community can be easily distilled by computing the cut position that corresponds to the minimum conductance value, and taking it as the cutoff point along this ranked list of nodes. We can now summarize the method to extract the emerged community. This method is called EC (Extract Community).

S1. Remove the nodes whose associated probability is 0 from the sorted node list **L**;

S2. Compute the conductance value of the community corresponding to each cut position;

S3. Take the community corresponding to the minimum conductance as the extracted one.

Here we consider an example illustrating EC operation when extracting the emerged community, which is shown as figure 4 and corresponds to the case in figure 1. In figure 4(a), the blue curve denotes that the conductance value varies with the cut position. The pink triangle indicates the cutoff position corresponding to the minimum conductance value, while the black inverted triangle denotes the cutoff position corresponding to that of the average probability $\varepsilon$ as cutoff value. In figure 4(b), the pink line denotes the cutoff value corresponding to the cut position that makes the conductance of the community to be minimal. It's obvious that, EC can very effectively extract the community, which has emerged by UC, and the structural similarity between the extracted community and the real one is now much higher, 0.9379.

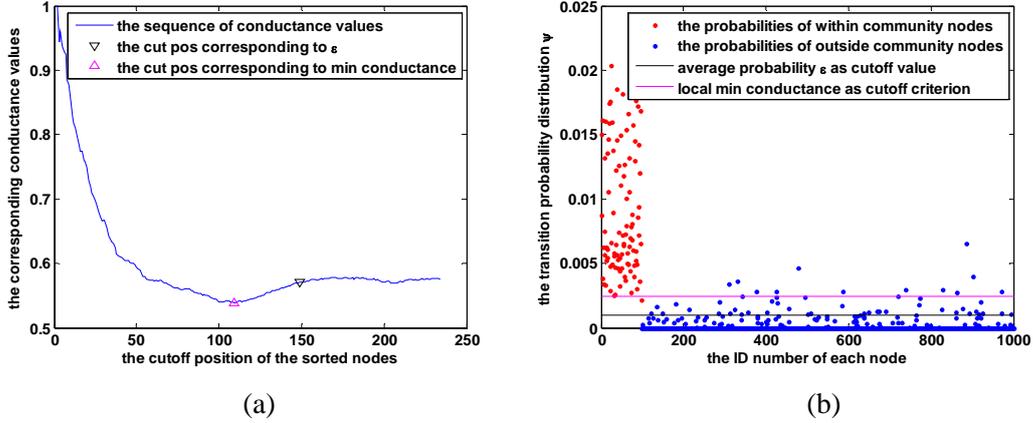

(a)   (b)

**Figure 4.** An illustration EC extraction of the emerged community, which comes from figure 1. **(a)** The conductance value for the extracted community as a function of the cutoff position of the sorted node sequence **L**. These nodes are ranked according to their probabilities in descending order. There are only 234 of 1000 nodes whose associated probabilities are greater than 0, thus the *x*-axis is just in the range of 1~234. **(b)** Extracting the emerged community by cutting the *l* step transition probability vector $\psi_s^l$ via two types of cutoff criterions.

**Proposition 2.** The time complexity of the EC is smaller than $O(dn^2)$, where *d* denotes the average degree of all the nodes.

*Proof.* It's obvious that S2 is the most computationally costly step in EC. We employ an incremental method to calculate the conductance value for each cut pos in the node list **L**. When the cut pos equals to 1, there is $S^1 = \{L(1)\}$ where $L(1)$ denotes the first node in the sorted node list $L$, $c_S^1 = d_{L(1)}$, $\text{Vol}(S^1) = d_{L(1)}$, and $\text{Vol}(V \setminus S^1) = m - \text{Vol}(S^1)$ where $m$ is the number of edges in the network. When the cut pos equals to $k$, there should be $S^k = S^{k-1} \cup \{L(k)\}$, $c_S^k = c_S^{k-1} + d_{L(k)} - 2*|S^k \cap N_{L(k)}|$ where $N_{L(k)}$ denotes the neighbor set of node $L(k)$, $\text{Vol}(S^k) = \text{Vol}(S^{k-1}) + d_{L(k)}$, and $\text{Vol}(V \setminus S^k) = m - \text{Vol}(S^k)$. It's obvious that, for each cut pos $k$, $S^k \cap N_{L(k)}$ is the most costly step, whose time complexity is $|S^k|*|N_{L(k)}| = k*d_{L(k)}$. Due to $k$ being placed in the range of $1 \leq k \leq k_{max}$ where $k_{max} << n$, the time complexity of EC should be $\sum_{k=1}^{k_{max}} k * d_{L(k)}$, which is much smaller than $O(dn^2)$.

**Proposition 3.** The time complexity of UEOC is $O(lKn^2)$, where $K$ denotes the number of communities.

*Proof.* From proposition 1 and proposition 2, it's obvious that the time to unfold and extract a community is $O((l+d)n^2)$. As there are $K$ communities in the cover, the time complexity of UEOC is $O((l+d)Kn^2)$. Since the average degree $d$ can be regarded as a constant as most complex networks are sparse graphs, it results that the time complexity of UEOC can also be given by $O(lKn^2)$.

There is a simple example to illustrate our UEOC method, which is shown as figure 5. As we can see, the UC subroutine induces two clear overlapping communities by calculating the transition probabilities. Meanwhile, the EC subroutine easily extracts these two communities by using our cut strategy. Furthermore, on this simple network this algorithm will attain convergence within five iterations.

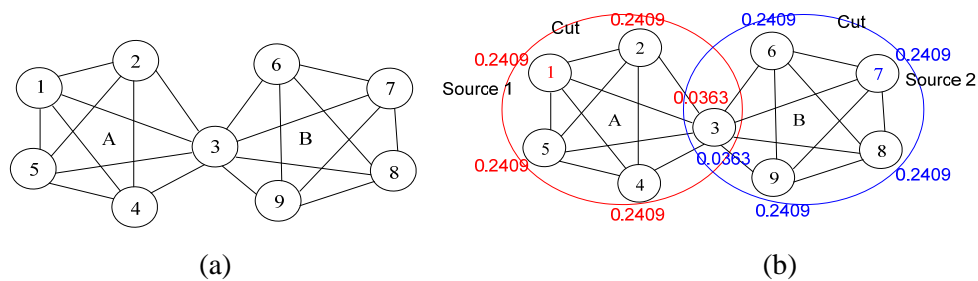

**Figure 5.** A simple example showing UEOC operation for the detection of overlapping communities. **(a)** A simple network with two overlapping communities. Node 3 is the overlapping node. **(b)** The result attained by the UEOC. The UC subroutine unfolds each source node's community by calculating the transition probabilities. The EC subroutine extracts each emerged community by using our cut strategy.

## 3. Experiments and evaluation

In order to evaluate the performance of algorithm UEOC, we tested it in benchmark computer-generated networks as well as on some widely used real-word networks. Two well-known algorithms CPM [3] and LFM [11] were selected to compare with UEOC. We conclude by analyzing the step number parameter $l$, which is defined in this algorithm.

There is only one parameter $l$ in UEOC. Here, we set $l = 20$ based on the experimental analysis in Sec. 3.3. Thus, the time complexity of UEOC can now be given by $O(Kn^2)$. For CPM, we set $k = 4$, which will return the best overall results [3]. For LFM, we set $\alpha = 1$, which is a natural choice, as it is the ratio of the internal degree to the total degree of the community [11].

All experiments are done on a single Dell Server (Intel(R) Xeon(R) CPU 5130 @

2.00GHz 2.00GHz processor with 4Gbytes of main memory), and the source code of the algorithms used here can be obtained from the authors.

*3.1. Computer-generated networks*

Here, we adopt two kinds of randomly generated synthetic networks (following both Newman model [1] and the Lancichinetti model [23]) with a known community structure, in order to evaluate the performance of the different algorithms.

There are various standard measures [27] that can be used to compare the known community structure of the benchmark and the one delivered by the algorithm. Unfortunately, most of these are not suitable for overlapping communities. The exception is a new variant of the Mutual Information measure, which has been widely used in the scientific area and was extended to handle overlapping communities [11]. We adopt this Normalized Mutual Information (NMI) as the accuracy measure in the following experiments.

*3.1.1. Benchmarks by Newman model.* The first type of synthetic networks employed here is that proposed by Newman at al. [1]. For these benchmarks, each graph consists of $n = 128$ vertices divided into four groups of 32 nodes. Each vertex has on average $z_{in}$ edges connecting it to members of the same group and $z_{out}$ edges to members of other groups, with $z_{in}$ and $z_{out}$ chosen so that the total expected degree $z_{in}+z_{out} = 16$. As $z_{out}$ is increased, starting from small initial values, the resulting graphs become more challenging to the community detection algorithms. Especially, when $z_{out}$ is greater than 8, meaning that the number of within-community edges is less than that of between-community edges for per vertex, the network doesn't have a community structure [1]. In figure 6(a), we show the NMI accuracy attained by each algorithm as a function of $z_{out}$. As we can see, our algorithm UEOC outperforms CPM and LFM in terms of NMI accuracy on these benchmarks.

Computational speed is another very important criterion to evaluate the performance of an algorithm. Time complexity analysis for UEOC has been covered under proposition 3 in Sec. 2.4 and refined by the second paragraph in Sec. 3. Nevertheless, here we show the actual running time of UEOC from an experimental angle, in order to further evaluate its efficiency.

Here we also address synthetic networks based on the Newman model [1]. In this case, each graph consists of $n = 40a$ vertices divided into forty groups of $a$ nodes. Each vertex has on average $z_{in} = 10$ edges connecting it to members of the same group and $z_{out} = 6$ edges to members of other groups. The only difference between the networks used here and the former ones is that, now $z_{out}$ is fixed while the community size $a$ is changeable. Figure 6(b)-(d) shows the actual running time of the UEOC and LFM. As we can see, when the community number $K$ is a constant, the square root of the running time by UEOC is nearly proportional to the number of nodes in the network, which is shown as figure 6(c). In the meantime, its efficiency is the same as that of the function $y=\mathbf{a}x^2+\mathbf{b}$, which is shown as figure 6(d). Thus, the experiment concurs with the analysis on the time complexity for UEOC, which is $O(Kn^2)$ when parameter $l$ is set at 20.

As far as we know, the worst-case computational complexity of LFM is $O(n^2\log n)$ [11], even though the efficiency of LFM seems to be better than UEOC on our test graphs, which is shown as figure 6(d). Meanwhile, the time complexity of CPM is non-polynomial even though it proves to be very efficient when applied to the real networks [3]. Thus, the efficiency (time complexity) of UEOC should be competitive with the efficiency of LFM, and higher than that of CPM.

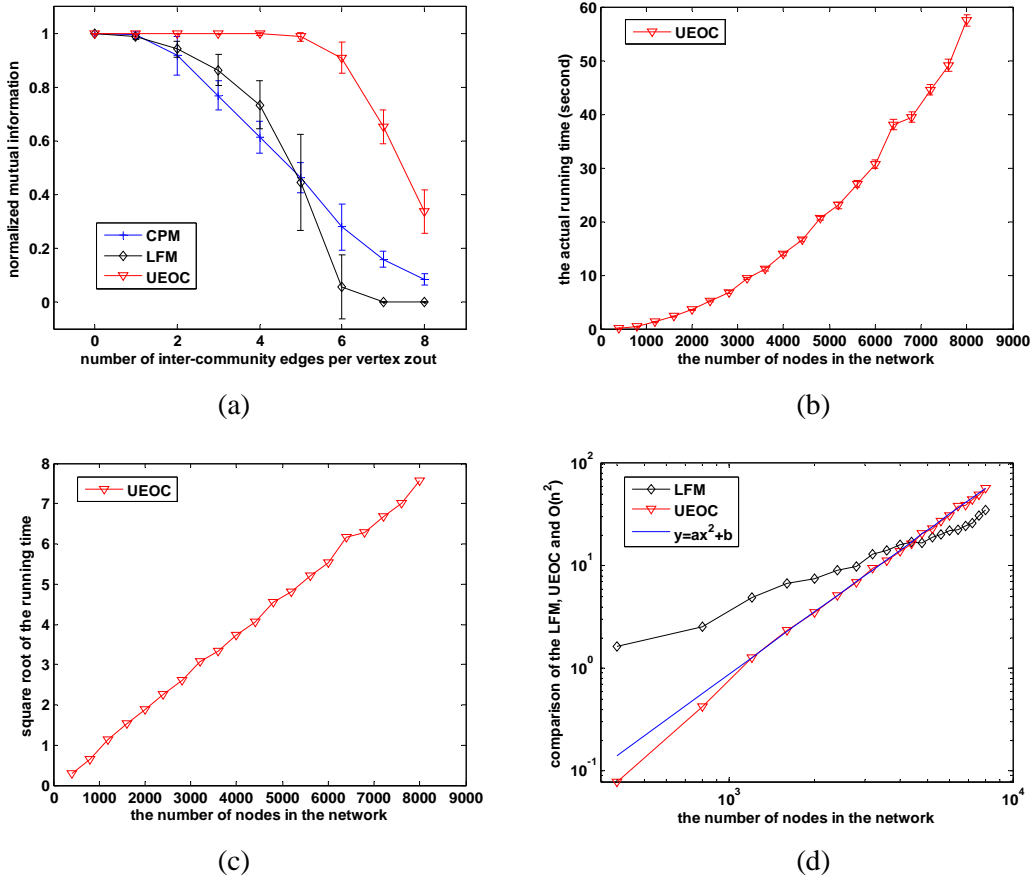

**Figure 6.** Testing the performance of UEOC on artificial networks under the Newman model. Error bars show the standard deviations estimated from 50 graphs. **(a)** Comparison of UEOC with CPM and LFM in terms of NMI accuracy. **(b)** Actual running time of UEOC as a function of the network scale. **(c)** Square root of the running time of UECO as a function of the network scale. **(d)** Efficiency of LFM, UECO and $O(n^2)$ under the log-log plots.

*3.1.2. Benchmarks by Lancichinetti model.* There are some basic statistical properties found in real networks, such as heterogeneous distributions of degree and community size, that are not found in benchmark networks based on Newman model. Accordingly, a new type of benchmark proposed by Lancichinetti et al. [23] is here adopted to further evaluate the accuracy of these algorithms. This class of benchmark networks not only has the property of heterogeneous distributions of degree and community size, but also can exhibit overlapping community structure.

Like the experiment designed by Lancichinetti et al. [23] to test CPM's ability to detect overlapping communities, the parameters setting for the Lancichinetti benchmarks are as follows. The network size $n$ is 1000, the minimum community size $c_{min}$ is set to either 10 or 20, the mixing parameter $\mu$ (each vertex shares a fraction $\mu$ of its edges with vertices in other communities) is set to either 0.1 or 0.3, the fraction of overlapping vertices ($o_n/n$) varies from 0 to 0.5 with interval 0.05. We keep the remaining parameters fixed: the average degree $d$ is 20, the maximum degree $d_{max}$ is 2.5*$d$, the maximum community size $c_{max}$ is 5*$c_{min}$, the number of communities, each overlapping vertex belongs to (denoted $o_m$), is 2, and the exponents of the power-law distribution of vertex degrees $\tau_1$ and community sizes $\tau_2$ are -2 and -1, respectively. This design space leads to four sets of benchmarks.

Figure 7 shows the results that compare UEOC with CPM and LFM in terms of NMI accuracy on the heterogeneous artificial networks with overlapping communities. As we can see, UEOC is most effective for networks with relatively big communities while CPM gives its best results for networks with small communities, and LFM has medium performance along both cases. Moreover, the quality of UEOC is relatively stationary with the increase of mixing parameter $\mu$, while that of the other two algorithms quickly declines in this situation. Thus, it's clear that our algorithm UEOC is competitive with, or better, than the other two algorithms on the Lancichinetti benchmarks.

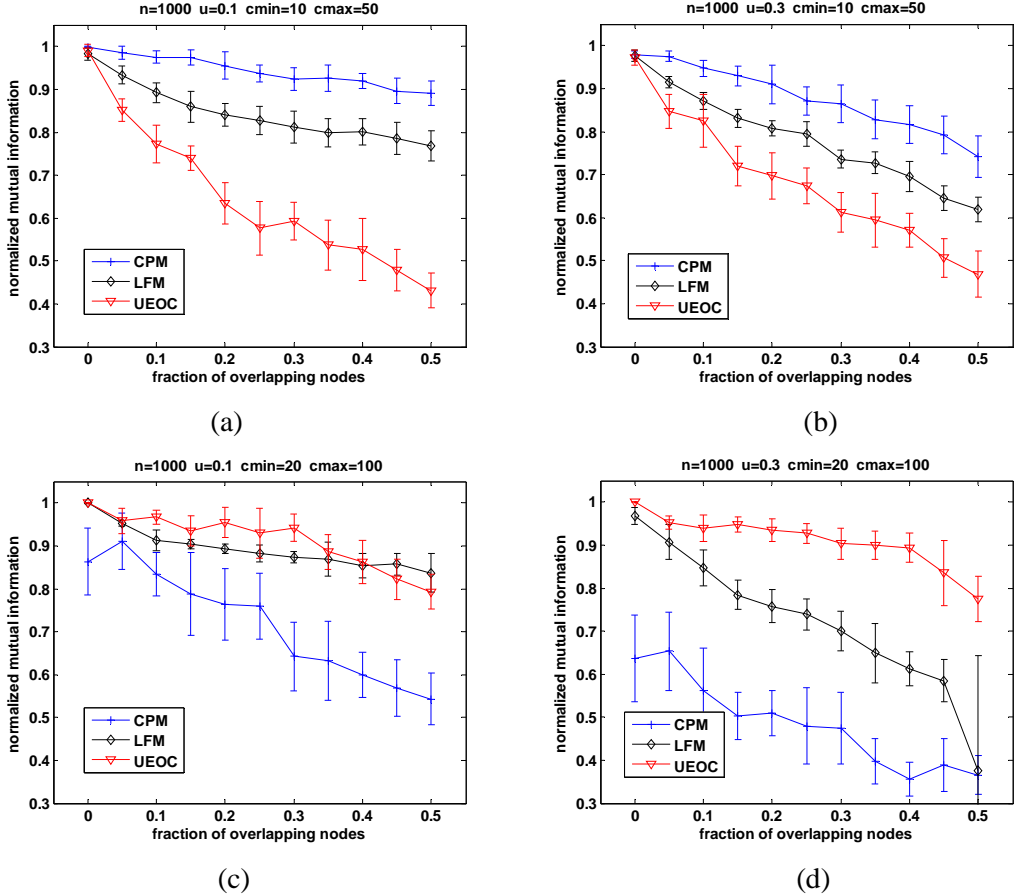

**Figure 7.** NMI accuracy of each algorithm as a function of the fraction of overlapping nodes. Error bars show the standard deviations estimated from 50 graphs. **(a)** Comparison on synthetic networks with small mixing parameter and small communities ($\mu = 0.1$, $c_{min} = 10$, $c_{max} = 50$). **(b)** Comparison on synthetic networks with big mixing parameter and small communities ($\mu = 0.3$, $c_{min} = 10$, $c_{max} = 50$). **(c)** Comparison on synthetic networks with small mixing parameter and big communities ($\mu = 0.1$, $c_{min} = 20$, $c_{max} = 100$). **(d)** Comparison on synthetic networks with big mixing parameter and big communities ($\mu = 0.3$, $c_{min} = 20$, $c_{max} = 100$).

*3.2. Real-world networks*
As real networks may have some different topological properties from the synthetic ones, we now consider several widely used real-world networks to further evaluate the performance of these algorithms. These networks are listed in table 1.

As the inherent community structure for real networks is usually unknown, we make use of two widely used quality measures for overlapping communities in order to evaluate the performance of the algorithms. The first one is the average conductance ($AC$), used by Leskovec et al. [22], which maps the average value of conductance for all the communities in a cover. The $AC$ is defined as (7), where $K$ denotes the number of communities, $C_i$ denotes the $i$-th community, and $\phi(S)$ denotes the conductance of a community $S$. The calculation of $\phi(S)$ has been already given in Sec. 2.4.

$$AC = \frac{1}{K}\sum_{i=1}^{K}\phi(C_i) \qquad (7)$$

The second one is a variant of the commonly used modularity ($Q$) metric [28], which is defined for overlapping communities by Shen et al. [5]. This extended modularity ($EQ$) is defined bellow (8), where $m$ denotes the number of total edges, $C_i$ denotes the $i$-th community, $O_v$ denotes the number of communities which node $v$ belongs to, $d_v$ denotes the degree of node $v$.

$$EQ = \frac{1}{2m}\sum_i\sum_{v,w\in C_i}\frac{1}{O_v O_w}\left[A_{vw} - \frac{d_v d_w}{2m}\right] \qquad (8)$$

It is noteworthy that, the quality of a cover is better when its $AC$-value is lower, whereas it is better when its $EQ$-value is higher.

Table 2 shows the results that compares UEOC with CPM and LFM in terms of both $AC$ and $EQ$ measures on the real-world networks described in table 1. As we can see, the $AC$ quality of UEOC is markedly better than that of CPM and LFM, while the $EQ$ quality of UEOC is also competitive with that of these two algorithms. To sum up, our algorithm is also very effective on real-word networks.

**Table 1.** Real-world networks used here.

| Networks | $|V|$ | $|E|$ | $1/\lambda_2$ | Descriptions |
|---|---|---|---|---|
| karate | 34 | 78 | 7.5602 | Zachary's karate club [29] |
| dolphin | 62 | 160 | 25.4027 | Dolphin social network [30] |
| polbooks | 105 | 441 | 26.4520 | Books about US politics [31] |
| football | 115 | 613 | 7.3097 | American College football [1] |
| email | 1,133 | 5,451 | 8.2563 | Emails of human interactions [32] |
| word | 7,207 | 31,784 | 6.8955 | Word semantic network [3] |

**Table 2.** The average result of 50 runs by UEOC, CPM and LFM on real networks.

|  | $AC$-values (the smaller the better) | | | $EQ$-values (the greater the better) | | |
|---|---|---|---|---|---|---|
|  | CPM | LFM | UEOC | CPM | LFM | UEOC |
| Karate | 0.6028 | **0.4750** | 0.5206 | 0.1147 | **0.3201** | 0.2648 |
| Dolphin | 0.3846 | 0.4494 | **0.3470** | 0.2908 | **0.3887** | 0.3846 |
| football | 0.3696 | 0.3895 | **0.2823** | 0.5593 | 0.5158 | **0.5996** |
| polbooks | 0.4093 | **0.2528** | 0.2749 | 0.4308 | **0.4744** | 0.4155 |
| email | 0.6937 | 0.7018 | **0.4627** | 0.2641 | 0.2282 | **0.3506** |
| word | 0.8089 | 0.7286 | **0.4553** | **0.1648** | 0.1624 | 0.1389 |

*3.3. Parameters analysis*

Sec. 2.3 shows that, if a network contains $k$ real communities, the convergence time of the UC should be a little longer than of the entering time in the $k$-*th* local mixing state (metastable state), which is $1/\lambda_{k+1}$. However, the actual community structure of most real-world networks is unknown. Thus, we adopt the entering time of the global mixing state ($1/\lambda_2$), which is greater than the entering time of the metastable state ($1/\lambda_{k+1}$), to coarsely evaluate the convergence time of the UC. As we can see from table 1, the $1/\lambda_2$ of all these real networks used in this paper is very small, meanwhile, they are found to be independent of the network scale. Thus, it shows that the UC will converge fast in the general case. Moreover, in order to further study the convergence characteristics of the UC, we also proceed to supply some quantitative analysis on the iteration number $l$ from an experimental angle.

In the UC, after each step, the transition probability vector $\psi_s^l$ will be updated and, thus, the ranking of all nodes will be changed according to the probability values incoming from source node $s$. As long as $\psi_s^l$ is stationary, or say, all members of the source community are put on the top of the sorted nodes sequence, it will be good enough for our purpose of unfolding a community.

Thus, the convergence of UC can be evaluated by considering the convergence of the transition probability vector, or that of the sorted node sequence. Figure 8 shows the convergence process of the UC with the increase of step number $l$. In figure 8(a), the $x$-axis refers to the $l$-value, and the $y$-axis refers to the difference between the two consecutive transition probability vectors, which is defined as the Euclidean distance between them. In figure 8(b), the $x$-axis still refers to the $l$-value, while the $y$-axis refers to the difference between the two consecutive sorted nodes lists $L_1$ and $L_2$, defined as nnz($L_1$–$L_2$), which counts the number of non-zeros of a given vector.

We have tested different $l$ values, in the range $1 \leq l \leq 20$, for all real networks mentioned in the paper. It's clear that the transition probability vector and the sorted node list can both converge well within 20 steps on each of the networks. Through these experiments, we also have found that UC in fact converges very quickly, thus its performance is insensitive to the choice of the parameter $l$ when it is greater than 20.

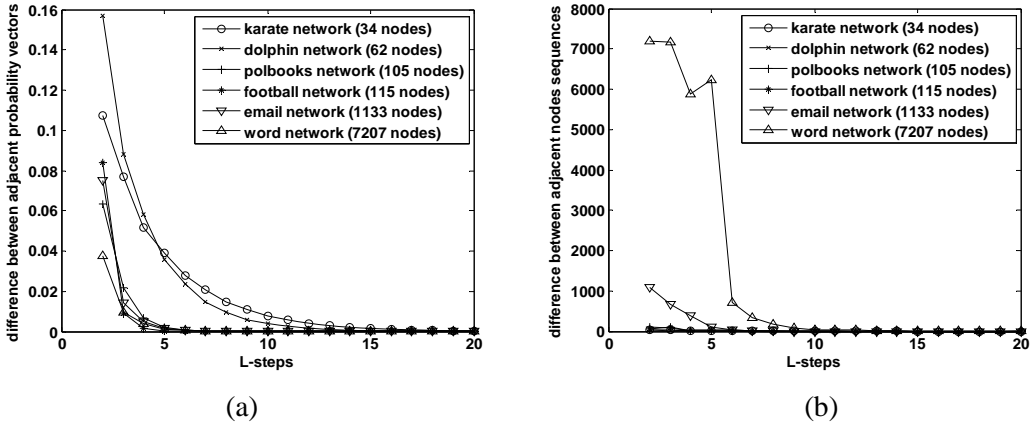

**Figure 8.** Convergence analysis of UC with the increase of parameter $l$. For each network, the node with maximum degree is selected as source node $s$ here. **(a)** Convergence process of the transition probability vector with the increase of step number $l$. **(b)** Convergence process of the sorted node sequence with the increase of step number $l$.

## 4. Discussion and conclusion

In this work, we have developed a new algorithm, called UEOC, for the identification of overlapping communities in complex networks. The key idea behind it rests on a Markov random walk method combined with a constraint strategy, based on which the UC phase can clearly unfold each natural community that overlaps. Thereafter, the community emerged by UC will be extracted from the entire network by the EC subroutine, which is an effective cutoff method based on a conductance function. Furthermore, the UEOC depends on only one parameter whose value can be easily set and requires no prior knowledge on the community structure, namely the number of communities.

We have tested this algorithm by using different types of networks. On Newman benchmarks, the accuracy of UEOC is higher than that of CPM and LFM. On Lancichinetti benchmarks, UEOC is most effective for networks with relatively big communities, and its clustering accuracy is relatively stationary with the increase of the mixing parameter. On real networks, the performance of our algorithm is in overall also better than that of the other two algorithms.

It is noteworthy that, there is also a similar approach so called Markov Cluster Algorithm (MCL) [14], which has been widely applied for discovering communities in graphs [33, 34]. The MCL and our method UEOC have in common that they both are based on the dynamic of Markov random walk, and detect the communities present in networks by changing and adjusting the Markov chain. However, they also have some important difference. The MCL simulates many random walks (or flows) on the network. By raising the transition probabilities to a certain power greater than one, it strengthens flow where it is already strong, and weakens flow where it is weak. By repeating the process an underlying community structure will gradually become visible. The process ends up with a number of regions with strong internal flow (communities), separated by 'dry' boundaries with hardly any flow. Different from the MCL, our method UEOC makes use of the distinction between the Markov random process on a network with community structure and that on its corresponding annealed network without communities. By preventing the walker (or flow) from escaping its own community, it makes the constrained Markov chain converge to a metastable state instead of the global mixing state, which then unfolds the real community structure of the network. Furthermore, the UEOC is a local community finding method that is able to detect the community of each node. Moreover, it can be also effectively used to discover overlapping communities from networks that hold them.

Other research, covering local methods for community detection, also harbors some potential for the detection of overlapping communities in complex networks. In 2002, Flake et al. [35] proposed an approach called Maximum Flow Communities (MFC). This method is based on the Max Flow-Min Cut theorem [36], stating that the maximum flow is identical to the minimum cut. Therefore, if you know the maximum flow between two points, you also know what edges you would have to remove to completely disconnect the same two points. The MFC has been successfully applied to the Web community mining area. However, this method is sensitive to the choice of its initial seed vertices, and its computational complexity, $O(mn\log(n^2/m))$, is significant. In 2004, Costa [37] presents a hub-based approach to community finding in complex networks. After identifying the network nodes with highest degree (the so-called hubs), the network is flooded with wavefronts of labels emanating from the hubs, accounting for the identification of the involved communities. This method is simple, efficient and proved to be effective for some social networks organized around hubs. However, it has the following drawback: the number of communities detected is arbitrarily preassigned and the algorithm neglects the possibility of having two hubs within the same community. In 2005, Bagrow et al. [38] proposed another efficient hub-based method for detecting local communities. This algorithm works by expanding an $l$ shell outward from some starting vertex $j$ and comparing the change in total emerging degree to some threshold $\alpha$. When this change is lower than a given value $\alpha$, the $l$ shell ceases to grow and all vertices covered by shells of a depth $\leq l$ are listed as members of vertex $j$'s community. This method is also suitable for the graphs organized around hubs, such as most social networks. However, it sensitive to the choice of its starting vertex $j$ and the setting of the threshold value $\alpha$. A potential compensating solution could be that, one would run the algorithm multiple times by using different starting vertices, and then achieve a group consensus as to which vertices belong to which communities. Different from these previous works, our algorithm UEOC is based on the Markov random walk on networks. It simulates the constrained probability flow starting from an arbitrary source node, and makes the local community of this node gradually appear during the iteration process. It is noteworthy that, our method is insensitive to the choice of its only one parameter (step number $l$), and it can effectively discover each node's local community for the more general networks.

Concluding, our future work can be laid as follows. In most real-word networks, there are often some marginal nodes, which will cause no energy change even they belong to several different communities. Generally speaking, if one does not comprehend the real meanings of the network structure, it will be very difficult for an algorithm to distinguish whether a specific marginal node is an overlapping node which plays an important role in the network and belongs to multi-communities, or an outlier which is just a isolated role in the network and belongs to none of these communities. However, it's very important for an algorithm to be able to identify both overlapping nodes and outliers in the real world. As far as we know, most of the current approaches including our method UEOC can only discover overlapping nodes, but cannot distinguish outliers. Thus, it will be promising to design a type of algorithm that can not only detect overlapping nodes, but also outliers. Therefore, we will take this direction in our future work so as to further improve the presented UEOC method.

Moreover, as we argued, the efficiency of UEOC is competitive with the efficiency of LFM and higher than that of CPM. Nevertheless, in order to deal with some large-scale networks such as WWW, Internet etc., it's still important to improve it by exploring some potential optimization angles. Thus, we also intend to further use UEOC in some applied research areas, such as biological networks analysis, Web community mining, etc., and try to uncover and interpret the significant overlapping community structure that one can expect to be found.

**Acknowledgments**
This work was supported by National Natural Science Foundation of China under Grant Nos. 60873149, 60973088, the National High-Tech Research and Development Plan of China under Grant No. 2006AA10Z245, the Open Project Program of the National Laboratory of Pattern Recognition, and the Erasmus Mundus Project of European Commission.